\documentclass[sigconf]{acmart}

\usepackage{colortbl}
\usepackage{pifont}
\usepackage{stfloats}
\usepackage{multicol}
\AtBeginDocument{%
  }

\setcopyright{acmlicensed}
\copyrightyear{2025}
\acmYear{2025}
\setcopyright{acmlicensed}
\acmConference[MM '25] {Proceedings of the 33rd ACM International Conference on Multimedia}{October 27--31, 2025}{Dublin, Ireland.}
\acmBooktitle{Proceedings of the 33rd ACM International Conference on Multimedia (MM '25), October 27--31, 2025, Dublin, Ireland}
\acmISBN{979-8-4007-2035-2/2025/10}
\acmDOI{	https://doi.org/10.1145/3746027.3755578}

\acmSubmissionID{4571}

\begin{document}

\title{ExplorAR: Assisting Older Adults to Learn Smartphone Apps through AR-powered Trial-and-Error with Interactive Guidance}

\author{Jiawei Li}
\email{jli526@connect.hkust-gz.edu.cn}
\orcid{0009-0000-6593-2958}
\affiliation{
  \institution{The Hong Kong University of Science and Technology (Guangzhou)}
  \city{Guangzhou}
  \country{China}
}

\author{Linjie Qiu}
\email{lqiu250@connect.hkust-gz.edu.cn}
\orcid{0009-0008-9204-5690}
\authornote{Contributed equally}
\affiliation{
  \institution{The Hong Kong University of Science and Technology (Guangzhou)}
  \city{Guangzhou}
  \country{China}
}

\author{Zhiqing Wu}
\email{zwu755@connect.hkust-gz.edu.cn}
\orcid{0009-0009-5417-5512}
\authornotemark[1]
\affiliation{
  \institution{The Hong Kong University of Science and Technology (Guangzhou)}
  \city{Guangzhou}
  \country{China}
}

\author{Qiongyan Chen}
\email{qchen580@connnect.hkust-gz.edu.cn}
\orcid{0009-0004-0893-7014}
\affiliation{
  \institution{The Hong Kong University of Science and Technology (Guangzhou)}
  \city{Guangzhou}
  \country{China}
}
\author{Ziyan Wang}
\email{lzwang082@connect.hkust-gz.edu.cn
}
\orcid{0000-0003-2465-2954}
\affiliation{
  \institution{The Hong Kong University of Science and Technology (Guangzhou)}
  \city{Guangzhou}
  \country{China}
}

\author{Mingming Fan}
\email{mingmingfan@ust.hk}
\orcid{0000-0002-0356-4712}
\authornote{Corresponding author}
\affiliation{
  \institution{The Hong Kong University of Science and Technology (Guangzhou)}
  \city{Guangzhou}
  \country{China}
  }
\affiliation{
  \institution{The Hong Kong University of Science and Technology}
  \city{Hong Kong SAR}
  \country{China}
}


\begin{abstract}

Older adults tend to encounter challenges when learning to use new smartphone apps due to age-related cognitive and physical changes. Compared to traditional support methods such as video tutorials, trial-and-error allows older adults to learn to use smartphone apps by making and correcting mistakes. However, it remains unknown how trial-and-error should be designed to empower older adults to use smartphone apps and how well it would work for older adults. Informed by the guidelines derived from prior work, we designed and implemented ExplorAR, an AR-based trial-and-error system that offers real-time and situated visual guidance in the augmented space around the smartphone to empower older adults to explore and correct mistakes independently. We conducted a user study with 18 older adults to compare ExplorAR with traditional video tutorials and a simplified version of ExplorAR. Results show that the AR-supported trial-and-error method enhanced older adults' learning experience by fostering deeper cognitive engagement and improving confidence in exploring unknown operations.
\end{abstract}



\maketitle

\section{INTRODUCTION}

The rapid advancement of digital technologies has led to online applications that simplify everyday tasks, such as accessing educational platforms, government services, and online banking~\cite{coursera, SHAREEF201117, EY}. While these innovations enhance convenience, they also pose challenges for individuals with limited digital literacy or cognitive ability~\cite{10.1145/3613904.3642423}. Older adults, in particular, often struggle with interface complexity due to age-related cognitive and physical changes, making technology less accessible to them~\cite{stossel2009evaluation, 10.1145/3613904.3642558}. Nearly 28\% of older adults report disabilities or health issues that hinder their use of digital tools, leading to lower engagement compared to their healthier counterparts~\cite{10.3389/frvir.2021.639718, app9173556, anderson2017technology, 10.1145/3613904.3642558}. To address these challenges, researchers have explored various support strategies. Traditional instructional materials and guidelines, such as video tutorials and step-by-step guides, have proven effective~\cite{Leung2012,fan2018guidelines}. More recently, trial-and-error learning has emerged as a promising approach, enabling older adults to explore applications independently with minimal cognitive load, aided by prompts and feedback~\cite{pang2021technology, Jin2022}.

Given the increasing need for accessible and engaging learning methods, augmented reality (AR) has emerged as a promising avenue for supporting older adults’ technology learning. In contrast to static instructional approaches, AR provides immersive, context-aware experiences by overlaying digital elements onto the real world~\cite{pandya2018taptag, pang2021technology, li2025remverse}. 
This interactivity can help reduce cognitive load and make learning more intuitive, particularly when paired with trial-and-error techniques and immediate feedback. 
Prior studies suggest that AR can support older adults in learning smartphone applications more effectively, particularly when combined with exploration to enhance confidence and skill acquisition~\cite{jin2024exploring}.
However, prior studies primarily focused on conceptual designs, while empirical evaluations of practical AR applications for supporting trial-and-error learning remain lacking~\cite{jin2024exploring}.
Moreover, previous research has insufficiently addressed the role of errors in trial-and-error learning within AR for older adults. These errors can interrupt the learning process, cause frustration, and ultimately undermine their motivation to continue engaging with the technology~\cite{barnard2013learning}.

To address these gaps, we propose ExplorAR, a novel trial-and-error approach that integrates AR to improve the learning experience for older adults. 
ExplorAR combines real-time error feedback and error-recovery support, not only providing real-time feedback when older adults make mistakes but also helping them recover from errors by identifying where they deviated from the correct interaction path. At the core of ExplorAR is ExplorTree, a hierarchical structure that captures screenshots and interactive elements at each step of the process, organizing them into ExplorNodes to guide users seamlessly.

To evaluate the effectiveness of ExplorAR in supporting older adults in learning and completing tasks on smartphones, we conducted a user study with 18 participants aged 60 and above. The study compared three instructional approaches: a conventional video tutorial, an AR-based step-by-step tutorial without trial-and-error interaction, and ExplorAR.
Our findings indicate that both AR-based methods enable participants to learn and complete tasks more effectively than the video tutorial. 
In addition, ExplorAR can boost participants' confidence in independently learning smartphone apps by enhancing the learning experience and reducing the fear of making mistakes. Furthermore, participants also saw strong potential in extending the trial-and-error design in ExplorAR to operations that are more complex and risky.
Drawing on findings from the empirical study, we offer three design implications aimed at improving trial-and-error learning experiences for older adults using AR to learn smartphone applications.


\section{BACKGROUND AND RELATED WORK}
\subsection{Challenges in Using Smartphone Applications Among Older Adults}

Although smartphones can potentially improve the daily lives of older adults by offering various applications tailored to their needs~\cite{kurniawan2008older}, they often face challenges in using smartphone applications, which can prevent them from fully reaping the benefits.

A primary challenge is cognitive decline, which affects memory, processing speed, and the ability to navigate new technologies. As cognitive functions decline, older adults may struggle to retain knowledge or recall how to use applications, making learning and using smartphones more difficult~\cite{rosenberg2009perceived}. 
Additionally, psychological factors such as technology anxiety and low self-confidence exacerbate these challenges. Many older adults find new technologies overwhelming and may avoid using them due to fear of failure or perceived complexity. This reluctance, combined with stereotypes about aging and technology, further limits their ability to engage with smartphone applications~\cite{zhu2024staying, chen2014gerontechnology}.

These barriers highlight the need for learning support to help older adults overcome cognitive and psychological challenges. This work aims to explore technical approaches to enhance the learning process of smartphone applications for older adults, ultimately fostering meaningful engagement with digital technologies.

\subsection{Research on Supporting Older Adults Independently Learning Mobile Applications}

Older adults typically use either dependent strategies (e.g., seeking help from family or friends)~\cite{bureau2010ministry,correa2015brokering} or independent ones (e.g., following instructions)~\cite{Leung2012} to learn mobile apps. As fewer than half of social support can help consistently~\cite{bureau2010ministry}, supporting independent learning is crucial for confident and autonomous app use.

Researchers have explored two independent learning methods: instructional materials (via video instructions and step-by-step) and trial-and-error feedback.
For instructional materials, Live View provided real-time interface guidance and instructional videos to help users learn apps individually~\cite{Leung2012}. Besides digital instructions, Conte et al.~\cite{Conte2019} developed embodiments of help invocation interactions to support older adults in navigating unfamiliar mobile apps. While prior studies indicated that older adults preferred using manuals over the trial-and-error method~\cite{Leung2012}, Pang et al.~\cite{pang2021technology} pointed out that older adults favored trial-and-error method owing to the more instinctive and immediate way to approach tasks. 
Older adults reported manuals overly complicated and often needed to make annotations or detailed notes to remember how to carry out specific tasks. The trial-and-error approach is beneficial for them to explore mobile apps independently within a low cognitive load when provided with a supportive environment. Synapse~\cite{Jin2022} created a safe environment for older adults to explore, learn, and correct mistakes, highlighting the importance of exploration in the learning process. 

Although the trial-and-error method has shown its potential in supporting older adults learning smartphone apps,  there is limited understanding of how trial-and-error method could be combined with instructional materials.
In this work, we took a step toward exploring how to enhance the learning experience for older adults when using smartphone applications by leveraging instructional materials in a trial-and-error method. 




\subsection{AR technology in Supporting Learning Experiences}
AR has been shown to enhance learning experiences and improve learning outcomes by offering an extended view, interactive surfaces, and visual cues \cite{jin2024exploring}.  
As Suzuki noted~\cite{Suzuki2022}, AR extends interaction beyond the tangible world by introducing dynamic, context-aware virtual interfaces that overlay physical screens. Researchers have innovatively broadened the AR interaction canvas in digital interfaces, CAVE systems, and smartphones~\cite{li2025interecon, Hartmann2020, Nishimoto2019}. Additionally, by seamlessly integrating virtual content into the real-world environment, interactions become more intuitive, as indicated by Reipschläger~\cite{Reipschl2020}. This extended display, along with an intuitive interaction method making AR technology have the potential to enhance the learning experiences of older adults, who often prefer larger screens and simpler interactions~\cite{pandya2018taptag, pang2021technology, li2023exploring}.

Providing rich visual cues is one of the advantages of AR technology in improving learning experiences. 
Klopfer \cite{klopfer2008augmented} demonstrated that AR's ability to deliver situational and context-rich information can simplify complex concepts and foster collaboration. Its effectiveness in improving learning experiences compared to conventional textbooks is further evidenced in specialized fields such as medical education, where AR greatly enhances understanding and retention of anatomical concepts \cite{moro2021virtual}.

In sum, researchers have demonstrated that AR can boost learning motivation and enhance long-term memory retention by presenting complex concepts and facilitating real-world simulations through interactive elements. However, most of them targeted young users and specific learning scenarios, not learning smartphone applications. Jin et al. took the first step in exploring how to adopt AR to support older adults in learning smartphones~\cite{jin2024exploring}.
However, the design and effectiveness of trial-and-error methods in AR remain understudied, especially compared to step-by-step guidance and video tutorials. Moreover, little is known about handling mistakes in such learning, which may hinder progress and reduce user engagement~\cite{barnard2013learning}.
Therefore, this study also sought to explore in what ways AR can assist older adults in learning smartphone applications.
\section{Design Considerations}
\label{Sec:Design Opportunities}
From the prior work, we derive three types of design considerations to make the process of learning smartphone applications by trial-and-error more accessible to older adults. These three components comprises: \textit{ 1) AR interaction}, \textit{2) instructional materials}, and \textit{3)learning methods}.


We focused on two key design considerations for the \textit{AR interaction}, which is aim to enhance the natural interaction and user experience towards learning process in the immersive environment: \textbf{(D1.1) enhancing situated interaction through AR self} and \textbf{(D1.2) representation and interaction via real hands.}  Prior studies have demonstrated situated and immersive interaction's potential to reduce the frustration inherent in learning \cite{sulter2022speakapp}, and create an environment that enhances situated learning and the transferability of knowledge \cite{morelot2021virtual}. Moreover, we emphasize interaction via real hands to ensure intuitive contact with virtual elements and promote comfort in older adults' motor skills \cite{bauer2020potential}. 

To create effective \textit{instructional materials}, 
it is important to \textbf{(D2.1) provide structured instructions in AR.} Older adults prefer learning through structured instructions that help them navigate complex processes without feeling overwhelmed \cite{Leung2012}.  
Additionally, two design considerations regarding to visualization of the instruction materials are essential: \textbf{(D2.2) enlarging font size and the whole screen while learning in AR} to improve easier screen reading \cite{jin2024exploring}, and \textbf{(D2.3) visualizing instructions by clear visual cues} to enhance the learning experience.

For the \textit{learning methods}, we \textbf{(D3.1) visualize the operation process} to help older adults enhance their memory of the operation process. We also \textbf{(D3.2) intergrate the trial-and-error methods and error recovery mechanism into immersive AR environment.} 
We aim to encourage older adults to explore and learn smartphone from their mistakes through trial-and-error approach.
Additionally, older adults feel less confident in exploring app use independently and fear making mistakes because they are experiencing a cognitive decline that they perceive in their daily lives \cite{vaportzis2017older}. So it is crucial to provide them with error recovery support and instant feedback.

\section{ExplorAR}

Drawing upon design considerations, we introduce ExplorAR, a novel trial-and-error method designed to support older adults in learning and exploring smartphone applications in an AR environment.
ExplorAR comprises ExplorNode, ExplorTree, and an end-user AR application, designed to support older adults in learning and exploring smartphone applications through a trial-and-error mechanism with situated visual cues \textbf{(D1.1, D1.2)}. 

The center of ExplorAR's functionality is ExplorTree \textbf{(D2.1)}, a hierarchical structure automatically constructed to associate smartphone page and interactive elements at each step of the operation process, represented as ExplorNodes. 
By leveraging ExplorTree, ExplorAR can visualize and record the operational processes undertaken \textbf{(D2.2, D2.3, D3.1)} through an end-user Hololens Application \textbf{(D2.1)}, providing error warnings along with error recovery instructions while exploring smartphone applications \textbf{(D3.2)}.



\begin{figure*}[tbh!]
    \centering
    \includegraphics[width=2\columnwidth]{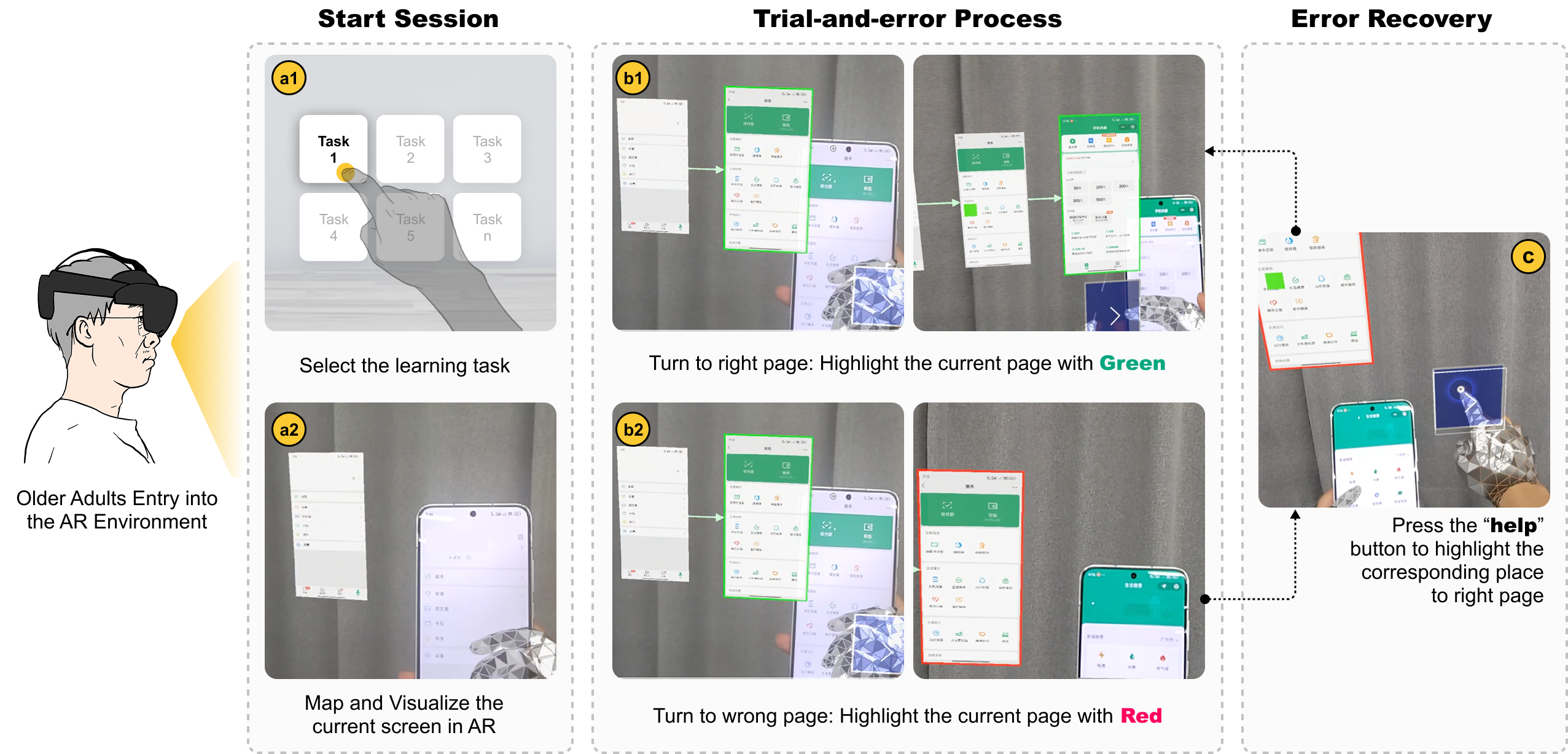}
    \caption{ExplorAR learning process in AR headset. (a1–a2) Users begin by selecting a task and mapping their current screen into the AR environment. (b1–b2) During the trial-and-error process, correct actions highlight the current page in green, while incorrect actions result in a red highlight. (c) If users encounter incorrect page or difficulties, pressing the "help" button provides visual guidance by highlighting the correct UI element leading to the right page.}
    \label{fig: explortreeBuild}
    \end{figure*}

As shown in Figure \ref{fig: explortreeBuild}, to illustrate how older adults can use ExplorAR, we follow an older adult, Bob, who is unable to recharge his phone bills online, and the ExplorAR usage process is as follows.

\begin{enumerate}
    \item Bob is using WeChat to recharge the mobile phone bill and he finds it difficult to locate the recharge entry among many services. To explore the correct recharge entry, Bob invokes the ExplorAR and retrieves the "Task1" tag.
    \item Bob then freely explores the platform, and when he navigates to the correct page, an enlarged digital screenshot with a green border is displayed on the AR device.
    \item If Bob navigates to the wrong page, the border of the current screenshot in AR will turn red as a warning.
    \item Bob can click the "Help" button when he needs assistance. Our system will provide both text and visual guidance for the navigation to the next correct page.
    \item Once the phone bill recharge process is completed, the entire usage history will be displayed. Bob can review it to identify mistakes he made and learn how to select the correct options.

\end{enumerate}




\begin{figure}[tbh!]
    \centering
    \includegraphics[width=1\columnwidth]{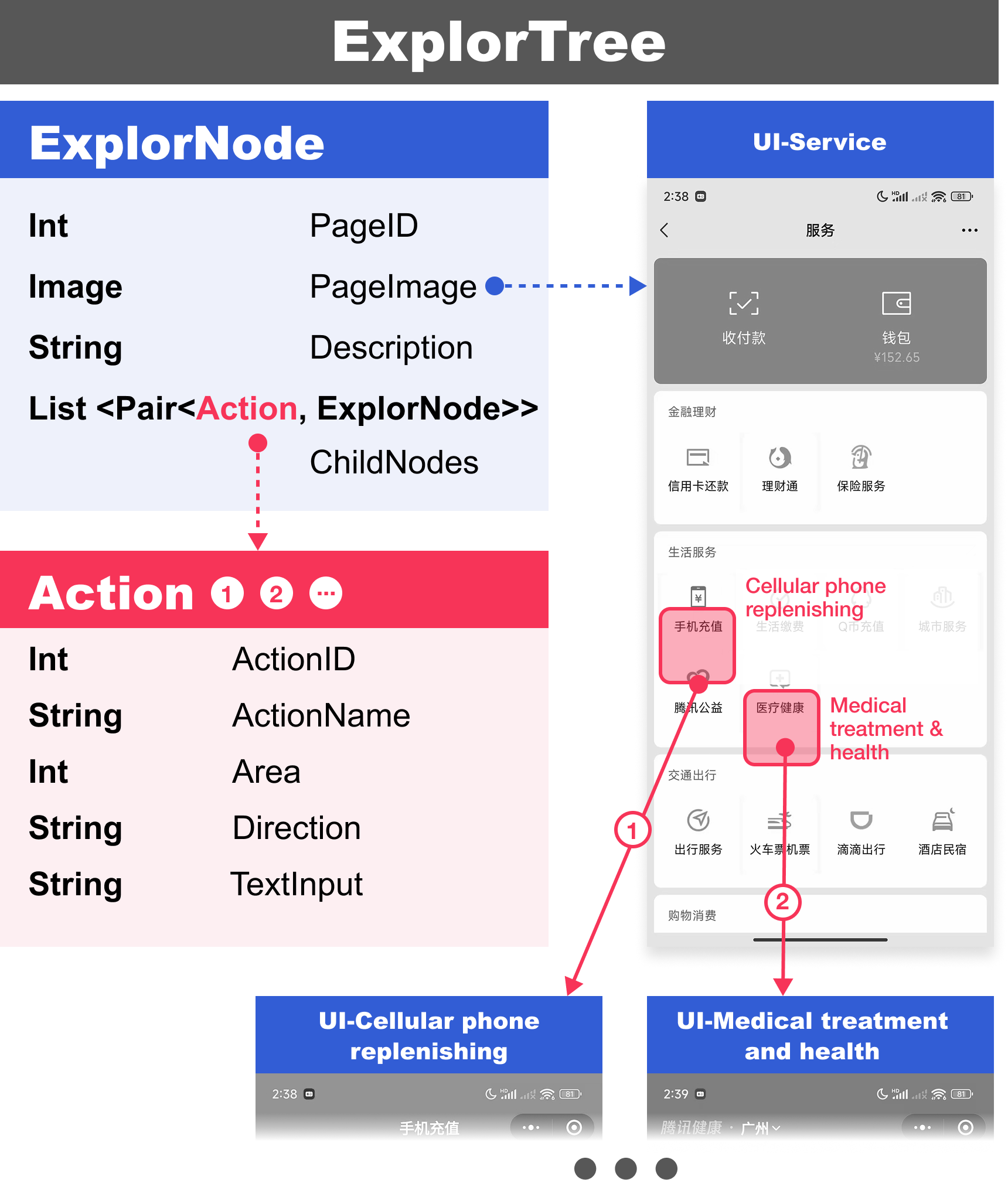}
    \caption{Overview of the ExplorTree structure. Each ExplorNode includes an ID, a PageImage, description, and a list of user actions that are defined by parameters including ActionID, ActionName, Area, Direction, and TextInput.}
    \label{fig: explortree}
    \end{figure}

 \begin{figure*}[tbh!]
    \centering
    \includegraphics[width=2\columnwidth]{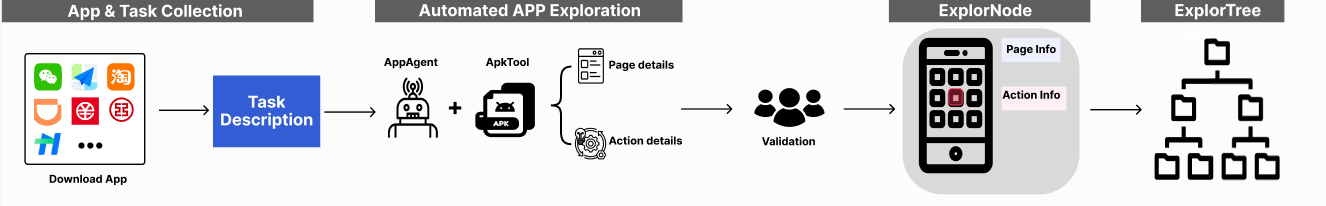}
    \caption{\textbf{ExplorNode \& ExplorTree data collection workflow. (1) Collecting apps and tasks. (2) Exploring and documenting app operation process information automatically by AI agents. (3) Manual validating and transforming peration process information into ExplorNode. (4) Construcing ExplorTree by connecting ExplorNode. }}
    \label{fig: datac}
    \end{figure*}

\subsection{ExplorNode \& ExplorTree: Operation Process Tree}

To create learning materials that support trial-and-error and error recovery. We automated the process of capturing screenshots and recording interactive elements at each step of operation process, representing these as ExplorNodes. Subsequently, we combined these ExplorNodes into the hierarchical structure of ExplorTree to represent the overall functionality, shown in Figure \ref{fig: explortree}. 
The representation of ExplorNode is: $$(PageID, PageImage, Description, <Action, ExplorNode>)$$
Each ExplorNode is composed of the following elements: 
\begin{itemize}
    \item $PageID$ is an identifier representing the specific page within the application.
    \item $PageImage$ is the screenshot of the user interface for the corresponding page, facilitating visual reference.
    \item $<Action, ExplorNode>$ is a list of mappings between user interactions (e.g., tapping or swiping a UI element) and the subsequent nodes, which depict the navigational flow to the next operational step.
\end{itemize}
 In detail, each $Action$ represents a user-triggered interaction required to transition from the current node to the next. $Actions$ are characterized by the following elements: (1) $ActionID$ is the identifier for the specific action. (2) $ActionName$ is the name of the action, such as "tap" or "swipe". (3) $Area$ is the UI coordinates where the action must occur. (4) $Direction$ specifies the swipe direction when the $ActionName$ equals "swipe". (5) $TextInput$ describes the text required to be entered as part of the action (\textit{e.g., }inputting a phone number or search term) when the $ActionName$ equals "
typing".

To implement ExplorNode and ExplorTree, we encourage LLM to interact with smartphone apps and record its interaction process as shown in Figure \ref{fig: datac}. We first identified and named each function, ensuring that each name is unique. 
For each function, we obtained the sequence of operations involved in its execution, including screenshots, step numbers, as well as the event required to move to the next step (\textit{e.g.,} taps, swipes). 
Subsequently, we used AppAgent \cite{yang2023appagent}, an LLM-based multimodal agent framework designed to operate smartphone applications to capture the interaction information for each function. Specifically, we provided the app name and task description to AppAgent and then demonstrated the function execution process to help AppAgent learn from the demo and generate documentation for the real-time screenshot showing the app’s interface and an XML file detailing the interactive elements in each step.
Subsequently, we validated the documentation manually and transformed it into json files to create standardized ExplorNode. 
Next, we constructed our ExplorTree by connecting the ExplorNode. 

\subsection{End-user AR Application to Support Guided Trial-and-error Learning}

\subsubsection{Screen and Operation Process Visualization}
As shown in Figure \ref{fig: explortreeBuild}a, ExplorAR enabled older adults to document their operation process and visualize the process in AR environment.
To achieve this, we first use the image recognition function from Vuforia API to identify the smartphone's current screen and then match the recognized screen with the PageImage attribute of ExplorNode to get the corresponding target node in the ExplorTree.
Then we present the screenshot of the matched ExplorNode by adding an extended screen in the AR environment.
Once a new screen is detected, its screenshot appears next to the previous interface's screenshot, with an arrow connecting them. This process sequentially presents the user's operation process within the AR environment.

\subsubsection{Trial-and-error and Error Recovery Design}
We design a \textit{trial-and-error mechanism} to assist older adults in exploring and learning the interaction of mobile applications as shown in Figure \ref{fig: explortreeBuild}b.
Specifically, when the AR camera recognizes a screen that matches one screenshot data in the ExplorTree, and the user clicks the \textbf{corresponding position to the right page} in the task, the current screenshot will be enlarged and outlined with a green border. If user clicks \textbf{wrong position to the page does not match the right page}, the screenshot will be outlined with a red border. Users receive a prompt only when they turn to the right page for the first time and will be prompted if they continue to click on the incorrect position to the wrong page, they are prompted when they complete a proper action for the first time. 
Additionally, if they navigate to an incorrect location or remain inactive for over 20 seconds, the system will provide prompt until they reach the correct location. 


To support trial-and-error, our system enables the \textit{error correction instructions} to guide older adults to seek help when receiving error operation warnings as shown in Figure \ref{fig: explortreeBuild}c. Specifically, the error correction instructions will only highlight the UI element leading to the next step in the current step's screenshot. It is worth mentioning that the instruction requires users to actively choose, they have the freedom to continue exploring or seek help based on their intention when they are unable to perform the correct operation.

\subsection{Implementation}
The ExplorAR was developed by Unity~\footnote{https://unity.com/cn/releases/editor/whats-new/2022.3.20\#notes} and is displayed through a Hololens 2 headset.  Additionally, we integrated the Mixed Reality Tool Kit (MRTK 3~\footnote{https://github.com/MixedRealityToolkit/MixedRealityToolkit-Unity}) to perform hand tracking and build user interface, ensuring seamless user interaction within the AR interface.


\section{Evaluation}
To examine how ExplorAR influences older adults’ learning processes, we conducted a user study with 18 participants and compared ExplorAR against two baseline methods: video-based tutorials (Video tutorial) and AR-based step-by-step instructions (AR instruction). The video tutorial simulates the real-life context in which older adults often learn smartphone apps independently through online videos. In contrast, AR instruction leverages augmented elements to provide interactive visual guidance, a promising yet underexplored approach in real-world settings \cite{jin2024exploring}. 
The comparison of these three methods is shown in the Table \ref{table:Comparsion}.

\subsection{Participants}
We recruited 18 older adults from the local older adult community center, all of whom self-reported as right-handed, with no motor impairments, and had normal vision that allowed them to participate in the study independently. They also have a basic knowledge of using smartphones. Among the 18 participants, 13 are female and 5 are male, with an average age of 65.6 years old (\textit{sd} = 2.91). 4 participants were using iPhones in their daily lives and the rest of the participants were all using Android phones. Each participant was compensated \$20 for their time.

We investigated the smartphone learning status of the participants. Among them, 6 participants preferred using online video to learn smartphone apps and 10 participants preferred asking for help from family members and people around them. Only 3 participants were willing to explore smartphone apps by themselves.


\subsection{Study Design}
To mitigate the learning effect between different conditions, we prerecorded tutorials with three different apps corresponding to three conditions.
We selected three apps that are commonly used by elderly people in their daily lives, covering finance, shopping, and public transportation.
In the choice of tutorial tasks, we made sure participants had never used them before the experiment. We fixed the order of the apps and assigned each participant to one of six combination condition status orders, following a balanced Latin square to mitigate the ordering effects.

\begin{table*}[]
\centering
\label{table:baseline}
\caption{Comparison of ExplorAR, AR instruction and video tutorial}
\begin{tabular}{cccccc}
\hline
                                                                 & Features                                                                                                   & Tutorial Format                                                           & \begin{tabular}[c]{@{}c@{}}Age-friendly \\ Design\end{tabular} & Equipment  & Presentation                                               \\ \hline
\rowcolor[HTML]{FFFFFF} 
ExplorAR (ours)                                                & \begin{tabular}[c]{@{}c@{}}Help older adults explore \\ and learn smartphone apps\end{tabular}             & \begin{tabular}[c]{@{}c@{}}Trial-and-error \&\\ Instructions\end{tabular} & \checkmark                                                            & AR         & \begin{tabular}[c]{@{}c@{}}Visual \& \\ Touch\end{tabular} \\
\rowcolor[HTML]{FFFFFF} 
\begin{tabular}[c]{@{}c@{}}AR instrucrion \end{tabular} & \cellcolor[HTML]{FFFFFF}\begin{tabular}[c]{@{}c@{}}Help older adults use \\ smartphone apps\end{tabular} & Instructions                                                              & \checkmark                                                             & AR         & Visual                                                     \\
\rowcolor[HTML]{FFFFFF} 
Video tutorial                                                           & \cellcolor[HTML]{FFFFFF}\begin{tabular}[c]{@{}c@{}}Help older adults use \\ smartphone apps\end{tabular} & Instructions                                                              & \ding{53}                                                            & Smartphone & \begin{tabular}[c]{@{}c@{}}Visual \& \\ Voice\end{tabular} \\ \hline
\end{tabular}
\label{table:Comparsion}
\end{table*}

\begin{figure*}[tbh!]
   \centering
   \includegraphics[width=2\columnwidth]{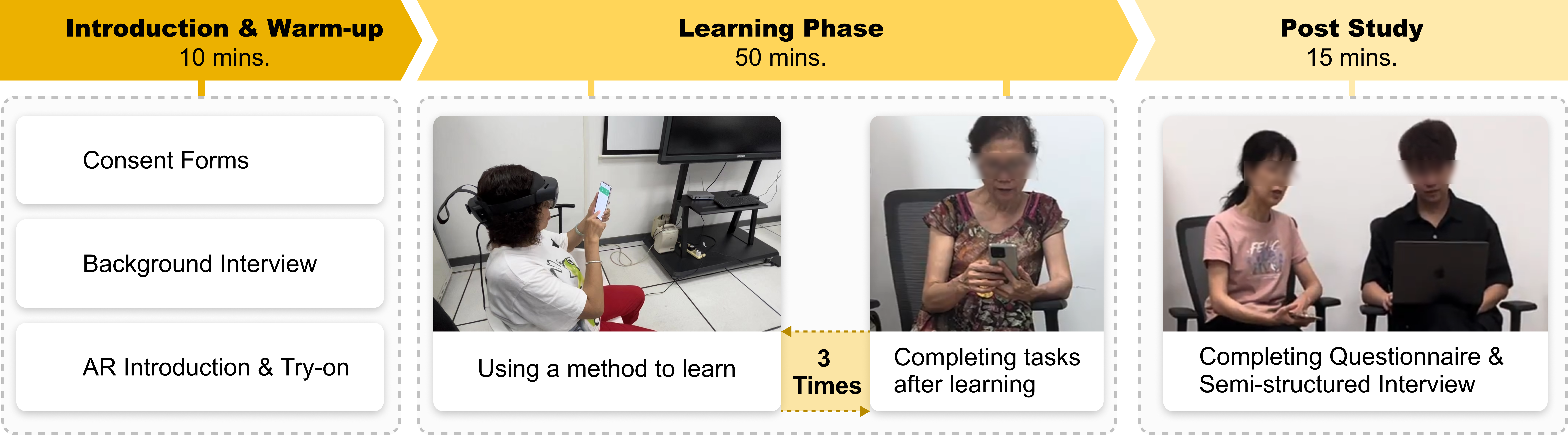}
  \caption{User study procedure. (a) Background interviews and AR introduction. (b) Use each method to learn and then complete tasks after learning. (c) Completed questionnaire and interviewed about experiences and perceptions in each learning method.}
  \label{fig: study}
 \end{figure*}

\subsection{Procedure}
The study lasted approximately 60 minutes. As shown in Figure \ref{fig: study}, before formal trials, participants filled in a questionnaire to collect their demographic and smartphone usage information. They were introduced to the experiment setup and signed a consent form. To help participants become familiar with AR, we asked participants to walk through the official HoloLens 2 tutorial to learn how to navigate the user interface with basic hand gestures. 

Before the experiment, we used the same case (“How to make an appointment using WeChat") for teaching and explained the three methods to participants.
The experiment has two sessions. In the first session,  participants were asked to learn three tasks, each with different methods (video tutorial, AR instructions, and ExplorAR) separately on three apps to avoid the learning effect from the previous condition. 
In the second session, to identify whether the condition of participants learned by the different methods would have any significant influence on their performance in learning smartphone apps, participants were asked to complete three similar tasks without support on these three apps in the last session, such as recharging mobile data instead of recharging phone credits. Participants had enough rest time to relax at the end of each session. 

After finishing the tasks, participants were asked to fill in another questionnaire after completing each session, where their subjective feedback on user experience using the UEQ-S questionnaire \cite{schrepp2017design}. 
In the end, we also conducted a semi-structured interview on their experience at the end of the study.

\subsection{Measurements}
For \textbf{objective measurements}, we aim to investigate and evaluate the effectiveness of the three methods in supporting older adults' learning smartphone apps. Specifically, we measured the time spent on each learning method, the time for completing each task, and the mistake counts they made during the task. Mistakes refer to clicking the wrong icon, which either directs the participant to the wrong page or leaves them stuck on the current page. 

For \textbf{subjective measurements}, we then included the result of the 7-point Likert scale UEQ-S questionnaire as a measure to evaluate the usability, conformability, and effectiveness of the method.


\subsection{Objective Result}
Utilizing the objective data from our study, we compared the learning effectiveness of the three learning methods in Table\ref{table:objective}.

\textbf{Learning Time: }
Participants spent significantly less time with the AR instructions during the learning process (\textit{avg} = 80.9s, \textit{sd} = 20.3), while spending more time with the ExplorAR (\textit{avg} = 109.2s, \textit{sd} = 44) and video tutorial (\textit{avg} = 112.9s \textit{sd} = 33.4). One reason that the AR-based method saves participants’ time is that they just need to follow the visual cues and click step by step, without extra thinking. 
Additionally, it is observed that although ExplorAR has a longer learning time, this was primarily because participants needed time to become accustomed to using the AR and to understand trial-and-error approach. Participants regarded this process as a novel experience and were willing to spend time on mastering it. 

\textbf{Task Completing Time: }
Our results revealed a significant difference in the time of independent task completion. Participants spent the least average time (\textit{avg} = 18.3s, \textit{sd} = 4.23) after learning from the ExplorAR method, while spending more time after learning from the AR instructions (\textit{avg} = 34.4s, \textit{sd} = 9.62) and video tutorials (\textit{avg} = 27.7s, \textit{sd} = 8.98). 
This result indicated that ExplorAR was the easiest to recall the steps while using video or AR instructions to learn the task resulted in lower memorability. 
For video tutorials, participants expressed difficulty in remembering the interaction process.
For AR instructions, an interesting point is that this method spent the least time during the learning phase also had the longest completion time and the highest error rate during the independent task phase.
In contrast, ExplorAR had the shortest completion time and the lowest error rate during the independent task phase. 

\textbf{Mistake Counts:} 
We counted the total mistake counts participants made while completing tasks independently after learning in each condition, and the result shows that participants made the fewest mistakes (n = 7) on the app, which they learned using ExplorAR, followed by the video tutorial (n = 13) and AR instructions (n = 18). This result indicated that ExplorAR generally achieved a higher
success rate after learning compared to the other two methods, where the basic mode with the lowest success rate. 
Additionally, participants praised the trial-and-error that enables them to explore by themselves, and hence they could remember the operation process. 
For the video tutorial mode, some of the participants expressed that it is similar to the basic mode, where they just followed the instructions, while another part of the participants would try to understand each step when watching the video. However, most of the participants forgot the steps after watching the video, half of them requested to watch the video again, and some of them requested to follow the steps of interaction by pausing the video.

\begin{table}[]
\centering
\caption{Objective results for each method.}
\begin{tabular}{cccc}
\hline
               & Learning Time    & \begin{tabular}[c]{@{}c@{}}Task Completing \\ Time\end{tabular} & \begin{tabular}[c]{@{}c@{}}Mistake\\ Counts\end{tabular} \\ \hline
AR instruction & 80.9s (sd=20.3)  & 34.4s (sd=9.62)                                                 & 18                                                       \\
Video tutorial & 112.9s (sd=33.4) & 27.7s (sd=8.98)                                                 & 13                                                       \\
ExplorAR       & 109.2s (sd=44.0) & 18.3s (sd=4.23)                                                 & 7                                                        \\ \hline
\end{tabular}
\label{table:objective}
\end{table}

\subsection{Subjective Result}
To present the subjective evaluation of participants, we first illustrate the quantitative results from the UEQ-S and then present qualitative results recorded in the post-stimulus survey.
\subsubsection{Overall Experience of ExplorAR}
\begin{figure}[tbh!]
    \centering
    \includegraphics[width=1\columnwidth]{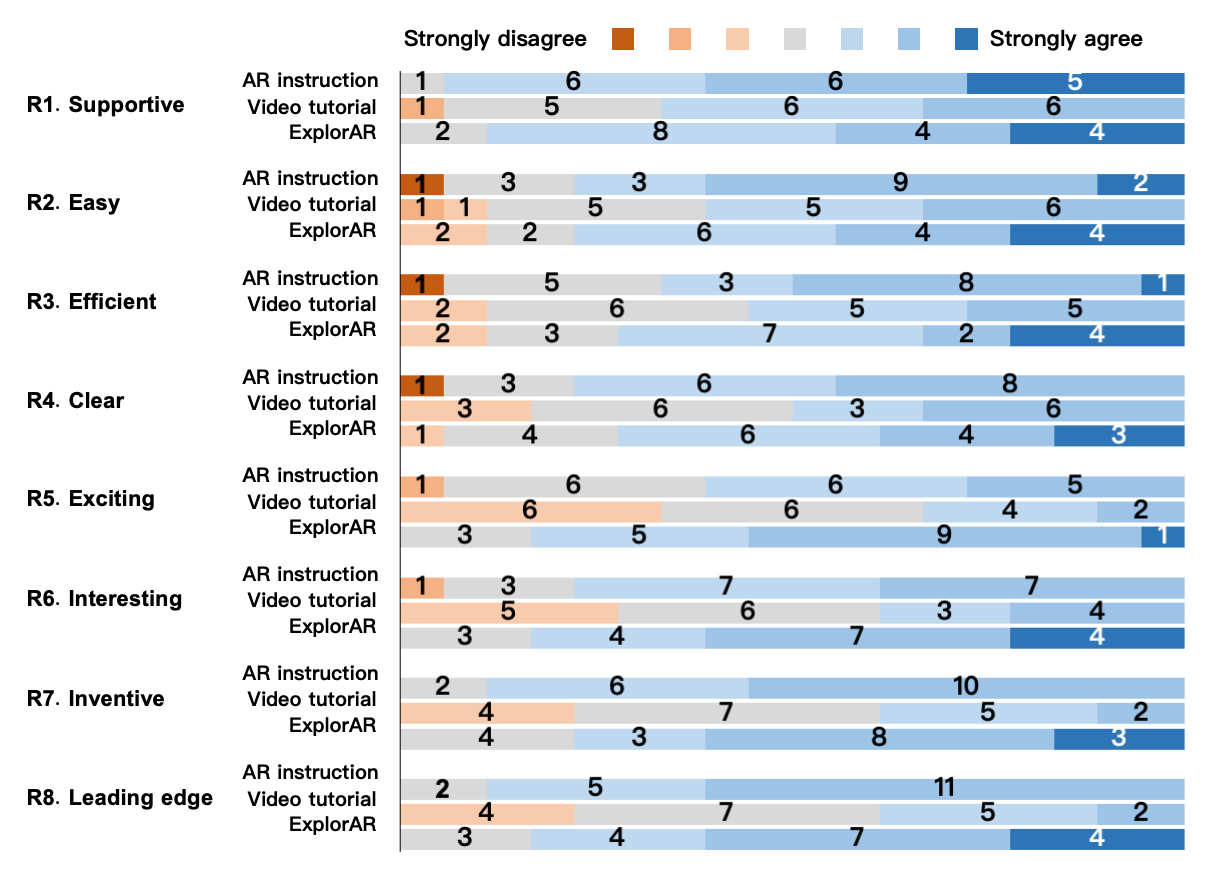}
    \caption{\textbf{UEQ-S questionaire rating scores for each method.}}
    \label{fig: ueq-short}
    \end{figure}

Figure \ref{fig: ueq-short} showed participants' rating scores for video, AR instructions, and ExplorAR in the UEQ-S questionnaire. 
By these ratings, we utilized the UEQ-S Table to calculate three short UEQ scales, including \textit{Pragmatic Quality}, \textit{Hedonic Quality}, and \textit{Overall}. 
Generally, participants rate higher scores for both two modes of the AR-based learning methods than for the recorded video (\textit{pragmatic quality} = 0.62, \textit{hedonic quality} = 0.28, and \textit{overall} = 0.45), while ExplorAR (\textit{pragmatic quality} = 1.31, \textit{hedonic quality} = 1.59, and \textit{overall} = 1.45) has a slight advantage over AR-based step-by-step (\textit{pragmatic quality} = 1.16, \textit{hedonic quality} = 1.21, and \textit{overall} = 1.18). 

In comparison with video and AR instructions, participants feel using ExplorAR to learn how to use smartphone apps is easier (\textit{avg} = 1.33, \textit{sd} = 1.28), clearer (\textit{avg} =  1.22, \textit{sd} = 1.17) and more supportive (\textit{avg} = 1.56, \textit{sd} = 0.98). Participants also appreciated ExplorAR, noting its significant assistance in enhancing the efficiency of learning (\textit{avg} = 1.17, \textit{sd} = 1.29). Participants found that using ExplorAR is more exciting (\textit{avg} = 1.44, \textit{sd} = 0.86) and interesting (\textit{avg} = 1.67, \textit{sd} = 1.03), they also rated higher inventive (\textit{avg} = 1.56, \textit{sd} = 1.04) and leading edge (\textit{avg} = 1.61, \textit{sd} = 1.09) scores. 

\subsubsection{Semi-structured Interview}
Participants reported that compared to video tutorials and AR instructions, ExplorAR enhanced their confidence in using and exploring smartphones, especially since they rarely took the initiative to learn new phone functions. While video tutorials were convenient because participants were accustomed to watching videos on their smartphones, ExplorAR offered additional reassurance and support, reducing the fear of making mistakes—particularly for illiterate older adults—through its trial-and-error approach. Still, some expressed concerns about the accuracy and authority of AI-driven methods. In terms of clarity, participants felt that AR-based approaches were generally clearer than video tutorials, which only provided verbal cues and often required replaying steps or searching for the correct position on the phone. They recommended salient visual or audio cues, as well as adapting AR interfaces for older adults with visual impairments. Most participants favored ExplorAR for reinforcing memory via correcting mistakes, reducing cognitive load, and potentially changing habits by drawing attention to incorrect operations; a few, however, preferred the AR instructions for its straightforward guidance or video tutorials for their familiarity. Lastly, participants saw strong potential in extending ExplorAR’s trial-and-error design to more complex operations, frequently updated applications (e.g., online shopping or service bookings), and high-risk tasks like making payments or revealing sensitive information.

\section{DISCUSSION}

\subsection{Trial-and-Error based Learning Method in AR}
In general, ExplorAR highlights the advantages of exploration-based learning, which encourages users to interactively engage with each step rather than passively follow instructional information. 
As demonstrated in our study, ExplorAR \textbf{fosters deeper cognitive processing} by allowing users to attempt, reflect, and adjust their actions during trial-and-error. 
Therefore, trial-and-error learning method is particularly beneficial for older adults, as it reduces the reliance on instructional information and instead promotes learning through action and correction, which echoes previous findings \cite{Jin2022}. 
We also find that exploration also \textbf{fosters confidence}, as users can safely explore and correct their mistakes without fear of making irreversible errors, which is particularly beneficial for older adults who may feel anxious about using unfamiliar digital tools \cite{barnard2013learning}. 
ExplorAR’s design thus illustrates that trial-and-error in learning can be a supportive method for encouraging older adults to learn and building their confidence. 

Our results also suggest that AR supports trial-and-error learning by the integration of visual, spatial, and interactive cues. In ExplorAR, AR allows users to see operations within the app context in real-time, which \textbf{reduces the need to switch attention between the app and external instructions}. By overlaying corrective feedback directly onto the interface, such as highlighting the correct action in green and incorrect actions in red, AR helps users understand and adjust their actions, thereby reinforcing correct operations. Additionally, AR’s ability to simulate real-world consequences in a controlled environment helps users \textbf{learn safe operations with less fear of real mistakes}, particularly in sensitive tasks like financial transactions. 


\subsection{Design Implications}
Drawing from our results, we identified four key design implications (DIs) for AR-based learning interaction techniques. These insights offer practical guidelines for future researchers and designers aiming to support older adults' learning in AR.

\textbf{DI1: Implement Personalized Learning Strategies.} Individual learning preferences and speeds vary significantly among older adults. Older adults should be allowed to adjust the level of guidance they receive, such as by choosing between ``Guided'' and “Independent Exploration” modes or setting task completion paces to enhance accessibility and user satisfaction \cite{mitzner2010older}. Adaptive learning strategies could further personalize the experience by adjusting difficulty based on user progress, providing additional support when needed and fostering independence.


\textbf{DI2: Prioritizing Visual Comfort in AR Design. }
Many participants experienced visual impairments or 3D discomfort when using AR headsets. To address this, the design of AR interfaces should prioritize visual comfort through larger font sizes, adjustable brightness, and customizable text contrast \cite{stearns2018design}. Incorporating adaptive focus settings can also accommodate varying levels of visual acuity \cite{stephanidis2021design}, ensuring that older adults can watch the content comfortably. Additionally, designing lightweight AR headsets can support a longer and more comfortable user experience \cite{chandana2023exploring}.

\textbf{DI3: Leveraging Multisensory Interaction.} Participants suggested that incorporating multisensory cues (\textit{e.g., }visual, auditory, and haptic feedback) can enhance their engagement and memory retention. In addition to visual and color-coded cues, the learning method should consider using spatial positioning and gentle sounds to indicate errors and correct actions.  This incorporation can help older adults reduce cognitive load and enable faster recall of learned steps. Therefore, future AR-based learning method designs should strategically employ multisensory interactions to make learning experiences more immersive, intuitive, and cognitively accessible.


\subsection{Generalizability}
\textbf{Generation to People Who Need to Learn Smartphones. }Although ExplorAR is initially designed for older adults to learn smartphone interaction, it could be adapted to other groups who may struggle with smartphone use (\textit{e.g., }people with limited prior exposure to technology, those with cognitive impairments, and younger users in emerging digital literacy programs \cite{blattgerste2019augmented, wei2024augmented}). 
With minimal modification, ExplorAR’s design and interaction flow can adapt to different groups' learning preferences by customizing error recovery support and interactive guidance.

\textbf{Generation to Other Digital Devices.} The design logic and key components of ExplorAR could aid accessibility research for other touchscreen devices. Its interaction paths and error recovery are applicable to similar touchscreen appliances (\textit{e.g.,} tablet computers and e-readers). Therefore, by slightly adjusting the logic and content to fit the device’s specific interfaces and functions, ExplorAR’s interactive AR model could provide a similarly immersive and accessible learning experience, assisting users in building transferable skills across various digital platforms.

\textbf{Wider Usage: Human Smartphone Usage Data Collection Tool. }The ExplorAR method could also serve as a tool for collecting human usage data on smartphone interactions. The trial-and-error logs, corrective feedback, and time-on-task data gathered during interactions could provide insights into common learning patterns and interaction preferences. This data could help refine future AR learning applications and inform user-centric design improvements for future research work.



\section{CONCLUSION}
To support older adults in learning and exploring smartphones, we introduced  ExplorAR, a guided trial-and-error method designed to foster an engaging, accessible, and effective learning experience in AR. By leveraging augmented reality, ExplorAR provides a safe environment where older adults can explore and correct mistakes through trial-and-error and visual guidance. Subsequently, we conducted an evaluation study with 18 older adults to compare the efficiency and effectiveness of this novel learning method against the video-based tutorial and AR-based step-by-step method. 
The results demonstrated that our proposed method significantly improved learning effectiveness and user experience, though it induced a longer learning time. We then further discussed the potential reasons and future design considerations for optimizing smartphone learning for older adults in AR. 

\section{Acknowledgement}
This work is partially supported by Guangzhou-HKUST(GZ) Joint Funding Project (No. 2024A03J0617), Guangzhou Higher Education Teaching Quality and Teaching Reform Project (No. 2024YBJG070),  Education Bureau of Guangzhou Municipality Funding Project  (No. 2024312152), Guangdong Provincial Key Lab of Integrated Communication, Sensing and Computation for Ubiquitous Internet of Things (No. 2023B1212010007), the Project of DEGP (No.2023KCXTD
042), and Artificial Intelligence Research and Learning Base of Urban Culture (No. 2023WZJD008). We would also thank Red Bird  MPhil Program at HKUST(GZ)  for their general support. 


\bibliographystyle{ACM-Reference-Format}
\bibliography{sample-base}


\begin{thebibliography}{44}


\ifx \showCODEN    \undefined \def \showCODEN     #1{\unskip}     \fi
\ifx \showISBNx    \undefined \def \showISBNx     #1{\unskip}     \fi
\ifx \showISBNxiii \undefined \def \showISBNxiii  #1{\unskip}     \fi
\ifx \showISSN     \undefined \def \showISSN      #1{\unskip}     \fi
\ifx \showLCCN     \undefined \def \showLCCN      #1{\unskip}     \fi
\ifx \shownote     \undefined \def \shownote      #1{#1}          \fi
\ifx \showarticletitle \undefined \def \showarticletitle #1{#1}   \fi
\ifx \showURL      \undefined \def \showURL       {\relax}        \fi
\providecommand\bibfield[2]{#2}
\providecommand\bibinfo[2]{#2}
\providecommand\natexlab[1]{#1}
\providecommand\showeprint[2][]{arXiv:#2}

\bibitem[Anderson and Perrin(2017)]%
        {anderson2017technology}
\bibfield{author}{\bibinfo{person}{Monica Anderson} {and} \bibinfo{person}{Andrew Perrin}.} \bibinfo{year}{2017}\natexlab{}.
\newblock \showarticletitle{Technology Use Among Seniors}.
\newblock \bibinfo{journal}{\emph{Washington, DC: Pew Research Center for Internet \& Technology}} (\bibinfo{year}{2017}).
\newblock


\bibitem[Barnard et~al\mbox{.}(2013)]%
        {barnard2013learning}
\bibfield{author}{\bibinfo{person}{Yvonne Barnard}, \bibinfo{person}{Mike~D Bradley}, \bibinfo{person}{Frances Hodgson}, {and} \bibinfo{person}{Ashley~D Lloyd}.} \bibinfo{year}{2013}\natexlab{}.
\newblock \showarticletitle{Learning to use new technologies by older adults: Perceived difficulties, experimentation behaviour and usability}.
\newblock \bibinfo{journal}{\emph{Computers in human behavior}} \bibinfo{volume}{29}, \bibinfo{number}{4} (\bibinfo{year}{2013}), \bibinfo{pages}{1715--1724}.
\newblock


\bibitem[Bauer and Andringa(2020)]%
        {bauer2020potential}
\bibfield{author}{\bibinfo{person}{Anna Cornelia~Maria Bauer} {and} \bibinfo{person}{Gerda Andringa}.} \bibinfo{year}{2020}\natexlab{}.
\newblock \showarticletitle{The Potential of Immersive Virtual Reality for Cognitive Training in Elderly}.
\newblock \bibinfo{journal}{\emph{Gerontology}} \bibinfo{volume}{66}, \bibinfo{number}{6} (\bibinfo{year}{2020}), \bibinfo{pages}{614--623}.
\newblock


\bibitem[Blattgerste et~al\mbox{.}(2019)]%
        {blattgerste2019augmented}
\bibfield{author}{\bibinfo{person}{Jonas Blattgerste}, \bibinfo{person}{Patrick Renner}, {and} \bibinfo{person}{Thies Pfeiffer}.} \bibinfo{year}{2019}\natexlab{}.
\newblock \showarticletitle{Augmented reality action assistance and learning for cognitively impaired people: a systematic literature review}. In \bibinfo{booktitle}{\emph{Proceedings of the 12th ACM international conference on pervasive technologies related to assistive environments}}. \bibinfo{pages}{270--279}.
\newblock


\bibitem[Bureau(2010)]%
        {bureau2010ministry}
\bibfield{author}{\bibinfo{person}{Statistics Bureau}.} \bibinfo{year}{2010}\natexlab{}.
\newblock \showarticletitle{Ministry of internal affairs and communications, Japan}.
\newblock \bibinfo{journal}{\emph{Annual Report on Current Population Estimates}} (\bibinfo{year}{2010}).
\newblock


\bibitem[Chen and Chan(2014)]%
        {chen2014gerontechnology}
\bibfield{author}{\bibinfo{person}{Ke Chen} {and} \bibinfo{person}{Alan Hoi~Shou Chan}.} \bibinfo{year}{2014}\natexlab{}.
\newblock \showarticletitle{Gerontechnology acceptance by elderly Hong Kong Chinese: a senior technology acceptance model (STAM)}.
\newblock \bibinfo{journal}{\emph{Ergonomics}} \bibinfo{volume}{57}, \bibinfo{number}{5} (\bibinfo{year}{2014}), \bibinfo{pages}{635--652}.
\newblock


\bibitem[Conte and Munteanu(2019)]%
        {Conte2019}
\bibfield{author}{\bibinfo{person}{Sho Conte} {and} \bibinfo{person}{Cosmin Munteanu}.} \bibinfo{year}{2019}\natexlab{}.
\newblock \showarticletitle{Help! I'm Stuck, and there's no F1 Key on My Tablet!}. In \bibinfo{booktitle}{\emph{Proceedings of the 21st International Conference on Human-Computer Interaction with Mobile Devices and Services}} (Taipei, Taiwan) \emph{(\bibinfo{series}{MobileHCI '19})}. \bibinfo{publisher}{Association for Computing Machinery}, \bibinfo{address}{New York, NY, USA}, Article \bibinfo{articleno}{10}, \bibinfo{numpages}{11}~pages.
\newblock
\showISBNx{9781450368254}
\href{https://doi.org/10.1145/3338286.3340121}{doi:\nolinkurl{10.1145/3338286.3340121}}


\bibitem[Correa et~al\mbox{.}(2015)]%
        {correa2015brokering}
\bibfield{author}{\bibinfo{person}{Teresa Correa}, \bibinfo{person}{Joseph~D Straubhaar}, \bibinfo{person}{Wenhong Chen}, {and} \bibinfo{person}{Jeremiah Spence}.} \bibinfo{year}{2015}\natexlab{}.
\newblock \showarticletitle{Brokering new technologies: The role of children in their parents’ usage of the internet}.
\newblock \bibinfo{journal}{\emph{New Media \& Society}} \bibinfo{volume}{17}, \bibinfo{number}{4} (\bibinfo{year}{2015}), \bibinfo{pages}{483--500}.
\newblock


\bibitem[{Coursera}(2012)]%
        {coursera}
\bibfield{author}{\bibinfo{person}{{Coursera}}.} \bibinfo{year}{2012}\natexlab{}.
\newblock \bibinfo{title}{Coursera}.
\newblock \bibinfo{howpublished}{\url{https://www.coursera.org/}}.
\newblock
\newblock
\shownote{Accessed: 2024-09-02}.


\bibitem[{Ernst \& Young}(2019)]%
        {EY}
\bibfield{author}{\bibinfo{person}{{Ernst \& Young}}.} \bibinfo{year}{2019}\natexlab{}.
\newblock \bibinfo{title}{Global FinTech Adoption Index 2019}.
\newblock \bibinfo{howpublished}{\url{https://assets.ey.com/content/dam/ey-sites/ey-com/en_gl/topics/banking-and-capital-markets/ey-global-fintech-adoption-index.pdf}}.
\newblock
\newblock
\shownote{Accessed: 2024-09-02}.


\bibitem[Fan and Truong(2018)]%
        {fan2018guidelines}
\bibfield{author}{\bibinfo{person}{Mingming Fan} {and} \bibinfo{person}{Khai~N Truong}.} \bibinfo{year}{2018}\natexlab{}.
\newblock \showarticletitle{Guidelines for creating senior-friendly product instructions}.
\newblock \bibinfo{journal}{\emph{ACM Transactions on Accessible Computing (TACCESS)}} \bibinfo{volume}{11}, \bibinfo{number}{2} (\bibinfo{year}{2018}), \bibinfo{pages}{1--35}.
\newblock


\bibitem[Fu et~al\mbox{.}(2024)]%
        {10.1145/3613904.3642423}
\bibfield{author}{\bibinfo{person}{Shihan Fu}, \bibinfo{person}{Jianhao Chen}, \bibinfo{person}{Emily Kuang}, {and} \bibinfo{person}{Mingming Fan}.} \bibinfo{year}{2024}\natexlab{}.
\newblock \showarticletitle{Bridging the Literacy Gap for Adults: Streaming and Engaging in Adult Literacy Education through Livestreaming}. In \bibinfo{booktitle}{\emph{Proceedings of the CHI Conference on Human Factors in Computing Systems}} (Honolulu, HI, USA) \emph{(\bibinfo{series}{CHI '24})}. \bibinfo{publisher}{Association for Computing Machinery}, \bibinfo{address}{New York, NY, USA}, Article \bibinfo{articleno}{658}, \bibinfo{numpages}{15}~pages.
\newblock
\showISBNx{9798400703300}
\href{https://doi.org/10.1145/3613904.3642423}{doi:\nolinkurl{10.1145/3613904.3642423}}


\bibitem[Hari~Chandana et~al\mbox{.}(2023)]%
        {chandana2023exploring}
\bibfield{author}{\bibinfo{person}{B. Hari~Chandana}, \bibinfo{person}{Nazeer Shaik}, {and} \bibinfo{person}{P. Chitralingappa}.} \bibinfo{year}{2023}\natexlab{}.
\newblock \showarticletitle{Exploring the Frontiers of User Experience Design: VR, AR, and the Future of Interaction}. In \bibinfo{booktitle}{\emph{2023 International Conference on Computer Science and Emerging Technologies (CSET)}}. \bibinfo{pages}{1--6}.
\newblock
\href{https://doi.org/10.1109/CSET58993.2023.10346724}{doi:\nolinkurl{10.1109/CSET58993.2023.10346724}}


\bibitem[Hartmann et~al\mbox{.}(2020)]%
        {Hartmann2020}
\bibfield{author}{\bibinfo{person}{Jeremy Hartmann}, \bibinfo{person}{Aakar Gupta}, {and} \bibinfo{person}{Daniel Vogel}.} \bibinfo{year}{2020}\natexlab{}.
\newblock \showarticletitle{Extend, Push, Pull: Smartphone Mediated Interaction in Spatial Augmented Reality via Intuitive Mode Switching}. In \bibinfo{booktitle}{\emph{Proceedings of the 2020 ACM Symposium on Spatial User Interaction}} (Virtual Event, Canada) \emph{(\bibinfo{series}{SUI '20})}. \bibinfo{publisher}{Association for Computing Machinery}, \bibinfo{address}{New York, NY, USA}, Article \bibinfo{articleno}{2}, \bibinfo{numpages}{10}~pages.
\newblock
\showISBNx{9781450379434}
\href{https://doi.org/10.1145/3385959.3418456}{doi:\nolinkurl{10.1145/3385959.3418456}}


\bibitem[Jin et~al\mbox{.}(2022)]%
        {Jin2022}
\bibfield{author}{\bibinfo{person}{Xiaofu Jin}, \bibinfo{person}{Xiaozhu Hu}, \bibinfo{person}{Xiaoying Wei}, {and} \bibinfo{person}{Mingming Fan}.} \bibinfo{year}{2022}\natexlab{}.
\newblock \showarticletitle{Synapse: Interactive Guidance by Demonstration with Trial-and-Error Support for Older Adults to Use Smartphone Apps}.
\newblock \bibinfo{journal}{\emph{Proc. ACM Interact. Mob. Wearable Ubiquitous Technol.}} \bibinfo{volume}{6}, \bibinfo{number}{3}, Article \bibinfo{articleno}{121} (\bibinfo{date}{Sept.} \bibinfo{year}{2022}), \bibinfo{numpages}{24}~pages.
\newblock
\href{https://doi.org/10.1145/3550321}{doi:\nolinkurl{10.1145/3550321}}


\bibitem[Jin et~al\mbox{.}(2024)]%
        {jin2024exploring}
\bibfield{author}{\bibinfo{person}{Xiaofu Jin}, \bibinfo{person}{Wai Tong}, \bibinfo{person}{Xiaoying Wei}, \bibinfo{person}{Xian Wang}, \bibinfo{person}{Emily Kuang}, \bibinfo{person}{Xiaoyu Mo}, \bibinfo{person}{Huamin Qu}, {and} \bibinfo{person}{Mingming Fan}.} \bibinfo{year}{2024}\natexlab{}.
\newblock \showarticletitle{Exploring the Opportunity of Augmented Reality (AR) in Supporting Older Adults to Explore and Learn Smartphone Applications} \emph{(\bibinfo{series}{CHI '24})}. \bibinfo{publisher}{Association for Computing Machinery}, \bibinfo{address}{New York, NY, USA}, Article \bibinfo{articleno}{21}, \bibinfo{numpages}{18}~pages.
\newblock
\showISBNx{9798400703300}
\href{https://doi.org/10.1145/3613904.3641901}{doi:\nolinkurl{10.1145/3613904.3641901}}


\bibitem[Klopfer(2008)]%
        {klopfer2008augmented}
\bibfield{author}{\bibinfo{person}{E. Klopfer}.} \bibinfo{year}{2008}\natexlab{}.
\newblock \bibinfo{booktitle}{\emph{Augmented Learning: Research and Design of Mobile Educational Games}}.
\newblock \bibinfo{publisher}{Penguin Random House LLC}.
\newblock
\showISBNx{9780262113151}
\showLCCN{2007032260}
\urldef\tempurl%
\url{https://books.google.com/books?id=I0kaFNaK704C}
\showURL{%
\tempurl}


\bibitem[Kurniawan(2008)]%
        {kurniawan2008older}
\bibfield{author}{\bibinfo{person}{Sri Kurniawan}.} \bibinfo{year}{2008}\natexlab{}.
\newblock \showarticletitle{Older people and mobile phones: A multi-method investigation}.
\newblock \bibinfo{journal}{\emph{International Journal of Human-Computer Studies}} \bibinfo{volume}{66}, \bibinfo{number}{12} (\bibinfo{year}{2008}), \bibinfo{pages}{889--901}.
\newblock


\bibitem[Lee et~al\mbox{.}(2019)]%
        {app9173556}
\bibfield{author}{\bibinfo{person}{Li~Na Lee}, \bibinfo{person}{Mi~Jeong Kim}, {and} \bibinfo{person}{Won~Ju Hwang}.} \bibinfo{year}{2019}\natexlab{}.
\newblock \showarticletitle{Potential of Augmented Reality and Virtual Reality Technologies to Promote Wellbeing in Older Adults}.
\newblock \bibinfo{journal}{\emph{Applied Sciences}} \bibinfo{volume}{9}, \bibinfo{number}{17} (\bibinfo{year}{2019}).
\newblock
\showISSN{2076-3417}
\href{https://doi.org/10.3390/app9173556}{doi:\nolinkurl{10.3390/app9173556}}


\bibitem[Leung et~al\mbox{.}(2012)]%
        {Leung2012}
\bibfield{author}{\bibinfo{person}{Rock Leung}, \bibinfo{person}{Charlotte Tang}, \bibinfo{person}{Shathel Haddad}, \bibinfo{person}{Joanna Mcgrenere}, \bibinfo{person}{Peter Graf}, {and} \bibinfo{person}{Vilia Ingriany}.} \bibinfo{year}{2012}\natexlab{}.
\newblock \showarticletitle{How Older Adults Learn to Use Mobile Devices: Survey and Field Investigations}.
\newblock \bibinfo{journal}{\emph{ACM Trans. Access. Comput.}} \bibinfo{volume}{4}, \bibinfo{number}{3}, Article \bibinfo{articleno}{11} (\bibinfo{date}{Dec.} \bibinfo{year}{2012}), \bibinfo{numpages}{33}~pages.
\newblock
\showISSN{1936-7228}
\href{https://doi.org/10.1145/2399193.2399195}{doi:\nolinkurl{10.1145/2399193.2399195}}


\bibitem[Li et~al\mbox{.}(2025a)]%
        {li2025remverse}
\bibfield{author}{\bibinfo{person}{Ruohao Li}, \bibinfo{person}{Jiawei Li}, \bibinfo{person}{Jia Sun}, \bibinfo{person}{Zhiqing Wu}, \bibinfo{person}{Zisu Li}, \bibinfo{person}{Ziyan Wang}, \bibinfo{person}{Ge~Lin Kan}, {and} \bibinfo{person}{Mingming Fan}.} \bibinfo{year}{2025}\natexlab{a}.
\newblock \showarticletitle{RemVerse: Supporting Reminiscence Activities for Older Adults through AI-Assisted Virtual Reality}.
\newblock \bibinfo{journal}{\emph{arXiv preprint arXiv:2507.13247}} (\bibinfo{year}{2025}).
\newblock


\bibitem[Li et~al\mbox{.}(2023)]%
        {li2023exploring}
\bibfield{author}{\bibinfo{person}{Zisu Li}, \bibinfo{person}{Li Feng}, \bibinfo{person}{Chen Liang}, \bibinfo{person}{Yuru Huang}, {and} \bibinfo{person}{Mingming Fan}.} \bibinfo{year}{2023}\natexlab{}.
\newblock \showarticletitle{Exploring the Opportunities of AR for Enriching Storytelling with Family Photos between Grandparents and Grandchildren}.
\newblock \bibinfo{journal}{\emph{Proceedings of the ACM on Interactive, Mobile, Wearable and Ubiquitous Technologies}} \bibinfo{volume}{7}, \bibinfo{number}{3} (\bibinfo{year}{2023}), \bibinfo{pages}{1--26}.
\newblock


\bibitem[Li et~al\mbox{.}(2025b)]%
        {li2025interecon}
\bibfield{author}{\bibinfo{person}{Zisu Li}, \bibinfo{person}{Jiawei Li}, \bibinfo{person}{Zeyu Xiong}, \bibinfo{person}{Shumeng Zhang}, \bibinfo{person}{Faraz Faruqi}, \bibinfo{person}{Stefanie Mueller}, \bibinfo{person}{Chen Liang}, \bibinfo{person}{Xiaojuan Ma}, {and} \bibinfo{person}{Mingming Fan}.} \bibinfo{year}{2025}\natexlab{b}.
\newblock \showarticletitle{InteRecon: Towards Reconstructing Interactivity of Personal Memorable Items in Mixed Reality}.
\newblock \bibinfo{journal}{\emph{arXiv preprint arXiv:2502.09973}} (\bibinfo{year}{2025}).
\newblock


\bibitem[Mitzner et~al\mbox{.}(2010)]%
        {mitzner2010older}
\bibfield{author}{\bibinfo{person}{Tracy~L Mitzner}, \bibinfo{person}{Julie~B Boron}, \bibinfo{person}{Cara~Bailey Fausset}, \bibinfo{person}{Anne~E Adams}, \bibinfo{person}{Neil Charness}, \bibinfo{person}{Sara~J Czaja}, \bibinfo{person}{Katinka Dijkstra}, \bibinfo{person}{Arthur~D Fisk}, \bibinfo{person}{Wendy~A Rogers}, {and} \bibinfo{person}{Joseph Sharit}.} \bibinfo{year}{2010}\natexlab{}.
\newblock \showarticletitle{Older adults talk technology: Technology usage and attitudes}.
\newblock \bibinfo{journal}{\emph{Computers in human behavior}} \bibinfo{volume}{26}, \bibinfo{number}{6} (\bibinfo{year}{2010}), \bibinfo{pages}{1710--1721}.
\newblock


\bibitem[Mor{\'e}lot et~al\mbox{.}(2021)]%
        {morelot2021virtual}
\bibfield{author}{\bibinfo{person}{Sarah Mor{\'e}lot}, \bibinfo{person}{Alain Garrigou}, \bibinfo{person}{Julie Dedieu}, {and} \bibinfo{person}{Bernard N'Kaoua}.} \bibinfo{year}{2021}\natexlab{}.
\newblock \showarticletitle{Virtual reality for fire safety training: Influence of immersion and sense of presence on conceptual and procedural acquisition}.
\newblock \bibinfo{journal}{\emph{Computers \& Education}}  \bibinfo{volume}{166} (\bibinfo{year}{2021}), \bibinfo{pages}{104145}.
\newblock


\bibitem[Moro et~al\mbox{.}(2021)]%
        {moro2021virtual}
\bibfield{author}{\bibinfo{person}{Christian Moro}, \bibinfo{person}{James Birt}, \bibinfo{person}{Zane Stromberga}, \bibinfo{person}{Charlotte Phelps}, \bibinfo{person}{Justin Clark}, \bibinfo{person}{Paul Glasziou}, {and} \bibinfo{person}{Anna~Mae Scott}.} \bibinfo{year}{2021}\natexlab{}.
\newblock \showarticletitle{Virtual and augmented reality enhancements to medical and science student physiology and anatomy test performance: A systematic review and meta-analysis}.
\newblock \bibinfo{journal}{\emph{Anatomical sciences education}} \bibinfo{volume}{14}, \bibinfo{number}{3} (\bibinfo{year}{2021}), \bibinfo{pages}{368--376}.
\newblock


\bibitem[Nishimoto and Johnson(2019)]%
        {Nishimoto2019}
\bibfield{author}{\bibinfo{person}{Arthur Nishimoto} {and} \bibinfo{person}{Andrew~E Johnson}.} \bibinfo{year}{2019}\natexlab{}.
\newblock \showarticletitle{Extending Virtual Reality Display Wall Environments Using Augmented Reality}. In \bibinfo{booktitle}{\emph{Symposium on Spatial User Interaction}} (New Orleans, LA, USA) \emph{(\bibinfo{series}{SUI '19})}. \bibinfo{publisher}{Association for Computing Machinery}, \bibinfo{address}{New York, NY, USA}, Article \bibinfo{articleno}{7}, \bibinfo{numpages}{5}~pages.
\newblock
\showISBNx{9781450369756}
\href{https://doi.org/10.1145/3357251.3357579}{doi:\nolinkurl{10.1145/3357251.3357579}}


\bibitem[Pandya and El-Glaly(2018)]%
        {pandya2018taptag}
\bibfield{author}{\bibinfo{person}{Shraddha Pandya} {and} \bibinfo{person}{Yasmine~N. El-Glaly}.} \bibinfo{year}{2018}\natexlab{}.
\newblock \showarticletitle{TapTag: Assistive Gestural Interactions in Social Media on Touchscreens for Older Adults}. In \bibinfo{booktitle}{\emph{Proceedings of the 20th ACM International Conference on Multimodal Interaction}} (Boulder, CO, USA) \emph{(\bibinfo{series}{ICMI '18})}. \bibinfo{publisher}{Association for Computing Machinery}, \bibinfo{address}{New York, NY, USA}, \bibinfo{pages}{244–252}.
\newblock
\showISBNx{9781450356923}
\href{https://doi.org/10.1145/3242969.3243003}{doi:\nolinkurl{10.1145/3242969.3243003}}


\bibitem[Pang et~al\mbox{.}(2021)]%
        {pang2021technology}
\bibfield{author}{\bibinfo{person}{Carolyn Pang}, \bibinfo{person}{Zhiqin Collin~Wang}, \bibinfo{person}{Joanna McGrenere}, \bibinfo{person}{Rock Leung}, \bibinfo{person}{Jiamin Dai}, {and} \bibinfo{person}{Karyn Moffatt}.} \bibinfo{year}{2021}\natexlab{}.
\newblock \showarticletitle{{Technology Adoption and Learning Preferences for Older Adults: Evolving Perceptions, Ongoing Challenges, and Emerging Design Opportunities}}. In \bibinfo{booktitle}{\emph{Proceedings of the 2021 CHI Conference on Human Factors in Computing Systems}} (Yokohama, Japan) \emph{(\bibinfo{series}{CHI '21})}. \bibinfo{publisher}{Association for Computing Machinery}, \bibinfo{address}{New York, NY, USA}, Article \bibinfo{articleno}{490}, \bibinfo{numpages}{13}~pages.
\newblock
\showISBNx{9781450380966}
\href{https://doi.org/10.1145/3411764.3445702}{doi:\nolinkurl{10.1145/3411764.3445702}}


\bibitem[Reipschl\"{a}ger et~al\mbox{.}(2020)]%
        {Reipschl2020}
\bibfield{author}{\bibinfo{person}{Patrick Reipschl\"{a}ger}, \bibinfo{person}{Severin Engert}, {and} \bibinfo{person}{Raimund Dachselt}.} \bibinfo{year}{2020}\natexlab{}.
\newblock \showarticletitle{Augmented Displays: Seamlessly Extending Interactive Surfaces With Head-Mounted Augmented Reality}. In \bibinfo{booktitle}{\emph{Extended Abstracts of the 2020 CHI Conference on Human Factors in Computing Systems}} (Honolulu, HI, USA) \emph{(\bibinfo{series}{CHI EA '20})}. \bibinfo{publisher}{Association for Computing Machinery}, \bibinfo{address}{New York, NY, USA}, \bibinfo{pages}{1–4}.
\newblock
\showISBNx{9781450368193}
\href{https://doi.org/10.1145/3334480.3383138}{doi:\nolinkurl{10.1145/3334480.3383138}}


\bibitem[Rosenberg et~al\mbox{.}(2009)]%
        {rosenberg2009perceived}
\bibfield{author}{\bibinfo{person}{Lena Rosenberg}, \bibinfo{person}{Anders Kottorp}, \bibinfo{person}{Bengt Winblad}, {and} \bibinfo{person}{Louise Nyg{\aa}rd}.} \bibinfo{year}{2009}\natexlab{}.
\newblock \showarticletitle{Perceived difficulty in everyday technology use among older adults with or without cognitive deficits}.
\newblock \bibinfo{journal}{\emph{Scandinavian journal of occupational therapy}} \bibinfo{volume}{16}, \bibinfo{number}{4} (\bibinfo{year}{2009}), \bibinfo{pages}{216--226}.
\newblock


\bibitem[Schrepp et~al\mbox{.}(2017)]%
        {schrepp2017design}
\bibfield{author}{\bibinfo{person}{Martin Schrepp}, \bibinfo{person}{Andreas Hinderks}, {and} \bibinfo{person}{Jörg Thomaschewski}.} \bibinfo{year}{2017}\natexlab{}.
\newblock \showarticletitle{Design and Evaluation of a Short Version of the User Experience Questionnaire (UEQ-S)}.
\newblock \bibinfo{journal}{\emph{International Journal of Interactive Multimedia and Artificial Intelligence}}  \bibinfo{volume}{4} (\bibinfo{date}{01} \bibinfo{year}{2017}), \bibinfo{pages}{103}.
\newblock
\href{https://doi.org/10.9781/ijimai.2017.09.001}{doi:\nolinkurl{10.9781/ijimai.2017.09.001}}


\bibitem[Seifert and Schlomann(2021)]%
        {10.3389/frvir.2021.639718}
\bibfield{author}{\bibinfo{person}{Alexander Seifert} {and} \bibinfo{person}{Anna Schlomann}.} \bibinfo{year}{2021}\natexlab{}.
\newblock \showarticletitle{The Use of Virtual and Augmented Reality by Older Adults: Potentials and Challenges}.
\newblock \bibinfo{journal}{\emph{Frontiers in Virtual Reality}}  \bibinfo{volume}{2} (\bibinfo{year}{2021}).
\newblock
\showISSN{2673-4192}
\href{https://doi.org/10.3389/frvir.2021.639718}{doi:\nolinkurl{10.3389/frvir.2021.639718}}


\bibitem[Shareef et~al\mbox{.}(2011)]%
        {SHAREEF201117}
\bibfield{author}{\bibinfo{person}{Mahmud~Akhter Shareef}, \bibinfo{person}{Vinod Kumar}, \bibinfo{person}{Uma Kumar}, {and} \bibinfo{person}{Yogesh~K. Dwivedi}.} \bibinfo{year}{2011}\natexlab{}.
\newblock \showarticletitle{e-Government Adoption Model (GAM): Differing service maturity levels}.
\newblock \bibinfo{journal}{\emph{Government Information Quarterly}} \bibinfo{volume}{28}, \bibinfo{number}{1} (\bibinfo{year}{2011}), \bibinfo{pages}{17--35}.
\newblock
\showISSN{0740-624X}
\href{https://doi.org/10.1016/j.giq.2010.05.006}{doi:\nolinkurl{10.1016/j.giq.2010.05.006}}


\bibitem[Stearns et~al\mbox{.}(2018)]%
        {stearns2018design}
\bibfield{author}{\bibinfo{person}{Lee Stearns}, \bibinfo{person}{Leah Findlater}, {and} \bibinfo{person}{Jon~E. Froehlich}.} \bibinfo{year}{2018}\natexlab{}.
\newblock \showarticletitle{Design of an Augmented Reality Magnification Aid for Low Vision Users}. In \bibinfo{booktitle}{\emph{Proceedings of the 20th International ACM SIGACCESS Conference on Computers and Accessibility}} (Galway, Ireland) \emph{(\bibinfo{series}{ASSETS '18})}. \bibinfo{publisher}{Association for Computing Machinery}, \bibinfo{address}{New York, NY, USA}, \bibinfo{pages}{28–39}.
\newblock
\showISBNx{9781450356503}
\href{https://doi.org/10.1145/3234695.3236361}{doi:\nolinkurl{10.1145/3234695.3236361}}


\bibitem[Stephanidis(2021)]%
        {stephanidis2021design}
\bibfield{author}{\bibinfo{person}{Constantine Stephanidis}.} \bibinfo{year}{2021}\natexlab{}.
\newblock \bibinfo{booktitle}{\emph{Design for All In Digital Technologies}}.
\newblock \bibinfo{publisher}{John Wiley \& Sons, Ltd}, Chapter~46, \bibinfo{pages}{1187--1215}.
\newblock
\showISBNx{9781119636113}
\href{https://doi.org/10.1002/9781119636113.ch46}{doi:\nolinkurl{10.1002/9781119636113.ch46}}


\bibitem[St{\"o}{\ss}el et~al\mbox{.}(2009)]%
        {stossel2009evaluation}
\bibfield{author}{\bibinfo{person}{Christian St{\"o}{\ss}el}, \bibinfo{person}{Hartmut Wandke}, {and} \bibinfo{person}{Lucienne Blessing}.} \bibinfo{year}{2009}\natexlab{}.
\newblock \showarticletitle{An evaluation of finger-gesture interaction on mobile devices for elderly users}.
\newblock \bibinfo{journal}{\emph{Prospektive Gestaltung von Mensch-Technik-Interaktion}}  \bibinfo{volume}{8} (\bibinfo{year}{2009}), \bibinfo{pages}{470--475}.
\newblock


\bibitem[S{\"u}lter et~al\mbox{.}(2022)]%
        {sulter2022speakapp}
\bibfield{author}{\bibinfo{person}{Robin~E S{\"u}lter}, \bibinfo{person}{Paul~E Ketelaar}, {and} \bibinfo{person}{Wolf-Gero Lange}.} \bibinfo{year}{2022}\natexlab{}.
\newblock \showarticletitle{SpeakApp-Kids! Virtual reality training to reduce fear of public speaking in children--A proof of concept}.
\newblock \bibinfo{journal}{\emph{Computers \& Education}}  \bibinfo{volume}{178} (\bibinfo{year}{2022}), \bibinfo{pages}{104384}.
\newblock


\bibitem[Suzuki et~al\mbox{.}(2022)]%
        {Suzuki2022}
\bibfield{author}{\bibinfo{person}{Ryo Suzuki}, \bibinfo{person}{Adnan Karim}, \bibinfo{person}{Tian Xia}, \bibinfo{person}{Hooman Hedayati}, {and} \bibinfo{person}{Nicolai Marquardt}.} \bibinfo{year}{2022}\natexlab{}.
\newblock \showarticletitle{Augmented Reality and Robotics: A Survey and Taxonomy for AR-enhanced Human-Robot Interaction and Robotic Interfaces}. In \bibinfo{booktitle}{\emph{Proceedings of the 2022 CHI Conference on Human Factors in Computing Systems}} (New Orleans, LA, USA) \emph{(\bibinfo{series}{CHI '22})}. \bibinfo{publisher}{Association for Computing Machinery}, \bibinfo{address}{New York, NY, USA}, Article \bibinfo{articleno}{553}, \bibinfo{numpages}{33}~pages.
\newblock
\showISBNx{9781450391573}
\href{https://doi.org/10.1145/3491102.3517719}{doi:\nolinkurl{10.1145/3491102.3517719}}


\bibitem[Vaportzis et~al\mbox{.}(2017)]%
        {vaportzis2017older}
\bibfield{author}{\bibinfo{person}{Eleftheria Vaportzis}, \bibinfo{person}{Maria Giatsi~Clausen}, {and} \bibinfo{person}{Alan~J. Gow}.} \bibinfo{year}{2017}\natexlab{}.
\newblock \showarticletitle{Older Adults Perceptions of Technology and Barriers to Interacting with Tablet Computers: A Focus Group Study}.
\newblock \bibinfo{journal}{\emph{Frontiers in Psychology}}  \bibinfo{volume}{Volume 8 - 2017} (\bibinfo{year}{2017}).
\newblock
\showISSN{1664-1078}
\href{https://doi.org/10.3389/fpsyg.2017.01687}{doi:\nolinkurl{10.3389/fpsyg.2017.01687}}


\bibitem[Wei et~al\mbox{.}(2024)]%
        {wei2024augmented}
\bibfield{author}{\bibinfo{person}{Qianjie Wei}, \bibinfo{person}{Jingling Zhang}, \bibinfo{person}{Pengqi Wang}, \bibinfo{person}{Xiaofu Jin}, {and} \bibinfo{person}{Mingming Fan}.} \bibinfo{year}{2024}\natexlab{}.
\newblock \showarticletitle{Augmented Library: Toward Enriching Physical Library Experience Using HMD-Based Augmented Reality}. In \bibinfo{booktitle}{\emph{Proceedings of the 17th International Symposium on Visual Information Communication and Interaction}}. \bibinfo{pages}{1--5}.
\newblock


\bibitem[Wu et~al\mbox{.}(2024)]%
        {10.1145/3613904.3642558}
\bibfield{author}{\bibinfo{person}{Zhiqing Wu}, \bibinfo{person}{Duotun Wang}, \bibinfo{person}{Shumeng Zhang}, \bibinfo{person}{Yuru Huang}, \bibinfo{person}{Zeyu Wang}, {and} \bibinfo{person}{Mingming Fan}.} \bibinfo{year}{2024}\natexlab{}.
\newblock \showarticletitle{Toward Making Virtual Reality (VR) More Inclusive for Older Adults: Investigating Aging Effect on Target Selection and Manipulation Tasks in VR}. In \bibinfo{booktitle}{\emph{Proceedings of the CHI Conference on Human Factors in Computing Systems}} (Honolulu, HI, USA) \emph{(\bibinfo{series}{CHI '24})}. \bibinfo{publisher}{Association for Computing Machinery}, \bibinfo{address}{New York, NY, USA}, Article \bibinfo{articleno}{24}, \bibinfo{numpages}{17}~pages.
\newblock
\showISBNx{9798400703300}
\href{https://doi.org/10.1145/3613904.3642558}{doi:\nolinkurl{10.1145/3613904.3642558}}


\bibitem[Zhang et~al\mbox{.}(2025)]%
        {yang2023appagent}
\bibfield{author}{\bibinfo{person}{Chi Zhang}, \bibinfo{person}{Zhao Yang}, \bibinfo{person}{Jiaxuan Liu}, \bibinfo{person}{Yanda Li}, \bibinfo{person}{Yucheng Han}, \bibinfo{person}{Xin Chen}, \bibinfo{person}{Zebiao Huang}, \bibinfo{person}{Bin Fu}, {and} \bibinfo{person}{Gang Yu}.} \bibinfo{year}{2025}\natexlab{}.
\newblock \showarticletitle{AppAgent: Multimodal Agents as Smartphone Users}. In \bibinfo{booktitle}{\emph{Proceedings of the 2025 CHI Conference on Human Factors in Computing Systems}} \emph{(\bibinfo{series}{CHI '25})}. \bibinfo{publisher}{Association for Computing Machinery}, \bibinfo{address}{New York, NY, USA}, Article \bibinfo{articleno}{70}, \bibinfo{numpages}{20}~pages.
\newblock
\showISBNx{9798400713941}
\href{https://doi.org/10.1145/3706598.3713600}{doi:\nolinkurl{10.1145/3706598.3713600}}


\bibitem[Zhu and Cheng(2024)]%
        {zhu2024staying}
\bibfield{author}{\bibinfo{person}{Xiaowen Zhu} {and} \bibinfo{person}{Xianping Cheng}.} \bibinfo{year}{2024}\natexlab{}.
\newblock \showarticletitle{Staying connected: smartphone acceptance and use level differences of older adults in China}.
\newblock \bibinfo{journal}{\emph{Universal Access in the Information Society}} \bibinfo{volume}{23}, \bibinfo{number}{1} (\bibinfo{year}{2024}), \bibinfo{pages}{203--212}.
\newblock


\end{thebibliography}



\end{document}